\documentclass[onecolumn,lineno,trackchanges,times]{aastex701}

\usepackage{amsmath}
\usepackage{amssymb}
\usepackage{bm}
\allowdisplaybreaks[4]

\newcommand{\DM}{\mathrm{DM}}

\newcommand{\IGM}{\mathrm{IGM}}

\newcommand{\host}{\mathrm{host}}
\newcommand{\ISM}{\mathrm{ISM}}

\newcommand{\LCDM}{$\Lambda$CDM}
\newcommand{\dd}{\mathrm{d}}

\begin{document}

\title{Probing the Bias of Large-Scale Structure with Unlocalized Fast Radio Bursts}

\author[0009-0003-4250-2199]{Yu-Tong Su}
\affiliation{School of Physics and Astronomy, Beijing Normal University, Beijing 100875, China}
\email{yutong.su@mail.bnu.edu.cn}

\author[0000-0002-8492-4408]{Zhengxiang Li}
\affiliation{School of Physics and Astronomy, Beijing Normal University, Beijing 100875, China}
\affiliation{Institute for Frontiers in Astronomy and Astrophysics, Beijing Normal University, Beijing 102206, China}
\email[show]{zxli918@bnu.edu.cn}

\begin{abstract}
Large-scale structure (LSS) and tracer bias provide a fundamental link between observable populations and the underlying matter distribution of the Universe. While galaxies are the standard tracers, recent studies have shown that transient sources such as gravitational-wave events can also probe LSS, even in the presence of significant localization uncertainties. Fast radio bursts (FRBs), with their cosmological distances and dispersion-measure (DM) information, offer a promising complementary probe of large-scale structure. However, the majority of FRBs lack precise localizations and redshift measurements, leading to substantial angular and radial uncertainties that smear the observable clustering signal.
In this work, we develop an end-to-end framework to infer the large-scale linear bias of unlocalized FRB populations using the isotropic two-point correlation function. Our method combines a Landy--Szalay estimator with noise-matched random catalogs, a forward model for localization-smearing based on Monte Carlo convolution, and likelihood-based inference with covariance matrices derived from lognormal mock catalogs.
Applying this pipeline to mock FRB samples at redshifts $z=0.3$, $0.5$, and $0.7$ with injected bias values $b=1.2$, $1.5$, and $2.0$, we find that the measured correlation functions consistently follow the localization-smeared theoretical prediction rather than the unsmeared one, confirming that positional uncertainties dominate the suppression of clustering. Despite significant realization scatter, the inferred bias posteriors generally recover the input values and preserve the ordering between different bias models. The separation is strongest at low redshift and degrades toward higher redshift, where low-bias populations become weakly constrained.
These results demonstrate that meaningful large-scale clustering information can be extracted from unlocalized FRB samples when localization effects are properly modeled, providing a viable pathway for future FRB-based LSS studies.
\end{abstract}

\keywords{fast radio bursts; large-scale structure; cosmology; dispersion measure; correlation functions}

\section{Introduction}
\label{sec:intro}

A central goal of observational cosmology is to map the large-scale structure (LSS) of the Universe and to relate the clustering of observable tracers to the underlying matter field. Two-point statistics provide one of the most direct characterizations of this structure, while tracer bias quantifies the statistical relation between the clustering of a given population and that of the underlying matter field on large scales \citep{Kaiser1984,DesjacquesJeongSchmidt2018}. In this sense, bias is not merely a nuisance parameter: it is also an astrophysical observable that connects source populations to their host environments and determines how effectively those populations can be used as cosmological tracers.

Galaxies and quasars are the canonical LSS tracers, the same statistical viewpoint can be extended to transient populations. In particular, recent work on gravitational-wave (GW) transients has shown that large-scale bias can be inferred even when source positions are only probabilistically localized, provided the localization uncertainty is forward modeled in the observable clustering signal \citep{Shao2022,Vijaykumar2023}. This motivates asking whether other classes of transients can also be used to trace LSS and constrain a meaningful large-scale bias.

Fast radio bursts (FRBs) are especially promising in this context. They are bright, millisecond-duration radio transients with extremely high brightness temperatures and large dispersion measures (DMs), implying extragalactic or cosmological propagation distances \citep{Lorimer2007,Thornton2013,Petroff2019,CordesChatterjee2019}. The rapid growth of FRB samples from wide-field instruments such as CHIME and ASKAP, together with future facilities including the SKA, has strengthened the case for FRBs as statistical probes of the intervening Universe. In particular, the second CHIME/FRB catalog already contains 4539 bursts from 3641 unique sources, making population-level cosmological studies increasingly realistic \citep{CHIMEFRB2026Cat2}.

A defining observable of an FRB is its dispersion measure,
\begin{equation}
\DM \equiv \int n_e\,\dd \ell ,
\end{equation}
which traces the free-electron column density along the line of sight and produces the characteristic $\nu^{-2}$ frequency-dependent arrival-time delay \citep{CordesChatterjee2019}. For an FRB at redshift $z$, the observed dispersion measure is commonly decomposed as
\begin{equation}
\DM_{\rm obs}(z) = \DM_{\ISM} + \DM_{\rm halo} + \DM_{\IGM}(z) + \frac{\DM_{\host}(z)}{1+z},
\label{eq:dm_budget}
\end{equation}
where $\DM_{\ISM}$ is the Milky Way interstellar-medium contribution, $\DM_{\rm halo}$ arises from the Galactic halo, $\DM_{\IGM}$ is the intergalactic component, and $\DM_{\host}$ accounts for the host galaxy and the immediate source environment. The factor $(1+z)^{-1}$ reflects the cosmological redshifting of the dispersive delay \citep{DengZhang2014,ProchaskaZheng2019,Macquart2020,WuZhangWang2022H0}. Because DM is sensitive to the ionized baryon distribution along the line of sight, FRBs naturally connect transient astrophysics to the LSS of the ionized Universe.

This connection has motivated a broad range of FRB-based cosmological applications. Localized FRBs with host-galaxy redshifts have already enabled a direct baryon census \citep{Macquart2020} and motivated constraints on cosmological parameters from DM--$z$ measurements \citep{Walters2018,Jaroszynski2019,WuZhangWang2022H0}. Complementary approaches include strongly lensed repeating FRBs as precision cosmological probes \citep{LiGao2018}, and the use of FRB DM statistics to study the thermal and ionization history of diffuse baryons, including He\,{\sc ii} reionization \citep{Zheng2014,Caleb2019,WeiGao2024HeII}. Recent reviews summarize these developments and related applications to fundamental physics and structure formation \citep{WuWang2024review,WangBaoLiuWei2026review}.

At the same time, the majority of FRBs detected by wide-field surveys do not yet have secure host-galaxy identifications or spectroscopic redshifts. This makes it natural to consider FRBs as a statistical LSS tracer even in the absence of one-to-one host associations. Existing studies have already explored angular clustering and cross-correlations between FRBs and external galaxy samples \citep{Shirasaki2017,RafieiRavandi2021LSS,SagaAlonso2024DMspace,Hussaini2025DMLSS}. These works demonstrate that FRBs carry measurable information about cosmic structure, but they also highlight a central challenge: for unlocalized events, coarse sky positions and DM-based radial estimates smear the reconstructed three-dimensional distribution and strongly suppress the observable clustering signal.

The goal of this paper is therefore to develop a practical framework for inferring the large-scale linear bias of \emph{unlocalized} FRB populations from the isotropic two-point correlation function. Our approach is inspired by analogous LSS analyses of GW transients and explicitly incorporates sky-localization and DM-based distance uncertainties into both the estimator and the forward model. In brief, we (i) construct correlation-function estimators with random catalogs matched to the \emph{observed} noise-convolved selection, (ii) compute a localization-smeared model correlation function by Monte Carlo convolution of the underlying matter clustering, and (iii) infer the FRB linear bias through likelihood-based comparison using covariance matrices estimated from ensembles of lognormal mock catalogs.

This paper is organized as follows. Section~\ref{sec:method} introduces the two-point correlation function estimator, the bias model, and the treatment of localization smearing. Section~\ref{sec:pipeline} describes the mock-catalog construction and inference pipeline. Section~\ref{sec:results} presents the recovered correlation functions and bias constraints. Section~\ref{sec:discussion} discusses the physical interpretation, limitations, and implications of the forecast.

\section{Method}
\label{sec:method}

\paragraph{Two-point correlation function and linear bias.}
For a number density field $n(\vec{x})=\bar{n}[1+\delta(\vec{x})]$, the two-point correlation function is defined as
\begin{equation}
\xi(r) \equiv \langle{\delta(\vec{x})\,\delta(\vec{x}+\vec{r})}\rangle,
\end{equation}
where angle brackets denote the ensemble average.
Under statistical homogeneity and isotropy, $\xi$ depends only on the separation distance $r=\lVert \vec{r}\rVert$.
On sufficiently large scales, a tracer population $X$ is commonly modeled as a biased tracer of the matter field,
\begin{equation}
\xi_X(r) = b_X^2\,\xi_m(r),
\end{equation}
where $\xi_X(r)$ is the tracer $X$'s 2PCF, $\xi_m(r)$ is the matter 2PCF, and $b_X$ is the (approximately scale-independent) linear bias.

\paragraph{Landy--Szalay estimator.}
We estimate $\xi(r)$ using the Landy--Szalay (LS) estimator \citep{LandySzalay1993},
\begin{equation}
\xi(r) = \left[
\frac{DD(r)}{N_D (N_D - 1)} - \frac{2DR(r)}{N_D N_R} + \frac{RR(r)}{N_R (N_R - 1)}
\right] \
\times \left[
\frac{RR(r)}{N_R (N_R - 1)}
\right]^{-1},
\end{equation}
where $DD(r)$, $DR(r)$, and $RR(r)$ are the pair counts in bins of separation $r$ between the data--data, data--random, and random--random catalogs.
$N_D$ and $N_R$ are the numbers of data and random points.
The random catalog is drawn to match the survey selection and geometry, and can contain many more points than the data catalog
to reduce shot noise in $DR(r)$ and $RR(r)$.

\paragraph{Localization smearing of the 2PCF.}
To estimate the 2PCF we require 3D positions of events and random points to compute pair separations.
For unlocalized FRBs, however, positions are uncertain in both sky coordinates and along the line of sight.
Such uncertainties distort the true spatial distribution and suppress clustering, particularly on small scales.

We account for this effect by forward-modeling a localization ``smearing'' of the underlying (true) 2PCF, following the GW LSS framework of
\citet{Vijaykumar2023} and related studies.
We model the positional uncertainty of each FRB as a Gaussian posterior in the 3D parameter space of (RA, Dec, distance).
For an FRB with true position $\mu = (\mathrm{RA}_0, \mathrm{Dec}_0, d_0)$, the observed positional probability density
$P(x)$ with $x = (\mathrm{RA}, \mathrm{Dec}, d)$ is
\begin{equation}
P(x) = \frac{1}{\sqrt{(2\pi)^3 |C|}} \exp\left[-\frac{1}{2}(x - \mu)^T C^{-1} (x - \mu)\right],
\end{equation}
where $C$ is the diagonal covariance matrix encoding uncertainties in RA ($\sigma_{\mathrm{RA}}$), Dec ($\sigma_{\mathrm{Dec}}$), and distance ($\sigma_d$).
This Gaussian approximation is typically adequate for a forecasting and validation pipeline when systematic biases are negligible.

The observed (noise-averaged) 2PCF corresponds to a convolution of the true correlation function with the localization kernels of event pairs.
Denoting the per-event displacement by $w$ and the pairwise displacement difference by $\Delta w \equiv w_2-w_1$, the smeared 2PCF can be written schematically as
\begin{equation}
\xi_{\mathrm{sm}}(r) = \int \dd^3 w_1 \int \dd^3 w_2 \, P(w_1)\,P(w_2)\,
\xi_{\mathrm{tr}}\!\left(\left| \bm{r} - (w_1-w_2)\right|\right),
\end{equation}
where $\xi_{\mathrm{tr}}$ is the true (unsmeared) correlation and $r$ is the observed pair separation.
In practice we work with the isotropic monopole and therefore spherically average over orientations of $\bm{r}$.

\paragraph{DM-redshift-distance uncertainty mapping.}
\label{sec:dmz}
For FRBs without secure host redshifts, the effective radial information is carried by the extragalactic DM.
For clustering, we require an \emph{effective line-of-sight comoving-distance uncertainty} $\sigma_r(z)$ that captures how DM scatter maps into radial smearing.
We propagate uncertainties through
\[
\sigma_{\DM}\ \rightarrow\ \sigma_z\ \rightarrow\ \sigma_r.
\]
We assume a flat \LCDM\ cosmology,
\begin{equation}
H(z)=H_0\sqrt{\Omega_m(1+z)^3+\Omega_\Lambda},\qquad
r(z)=\int_0^z \frac{c}{H(z')}\,\dd z'.
\end{equation}

\paragraph{Mean DM--$z$ relation.}
Starting from the DM budget in Eq.~\eqref{eq:dm_budget}, we isolate an effective $\DM_{\IGM}$ and adopt a mean $\DM$--$z$ relation.
A physically-motivated expression is \citep[e.g.,][]{DengZhang2014,Macquart2020}
\begin{equation}
\langle \DM_{\IGM}(z)\rangle =
\frac{3cH_0\Omega_b}{8\pi G m_p}
\int_0^z
\frac{f_{\IGM}(z')\,f_e(z')\,(1+z')}{E(z')}\,\dd z',
\qquad E(z)\equiv \frac{H(z)}{H_0}.
\end{equation}
For the purpose of an \emph{effective} smearing scale in a forecasting pipeline, we use the widely adopted pocket approximation motivated by the Macquart relation,
\begin{equation}
\langle \DM_{\IGM}\rangle \simeq 1000\,z\ \ (\mathrm{pc\ cm^{-3}}),
\qquad \frac{\dd \langle \DM\rangle}{\dd z}\simeq 1000\ \ (\mathrm{pc\ cm^{-3}}).
\end{equation}
(Using the full integral form above instead of the linear mapping changes $\sigma_r$ by a subdominant amount at the redshifts of interest for our forecasts.)

\paragraph{DM uncertainty budget.}
We treat the DM measurement error as negligible and model the dominant uncertainty as astrophysical scatter.
We add contributions in quadrature:
\begin{equation}
\sigma_{\DM}^2(z)=
\sigma_{\IGM}^2(z)+\sigma_{\host}^2(z)+\sigma_{\rm halo}^2+\sigma_{\ISM}^2 .
\end{equation}
The host contribution is assumed to scale as
\begin{equation}
\sigma_{\host}(z)=\frac{\sigma_{\host,0}}{1+z}.
\end{equation}
The IGM scatter is parameterized as a redshift-dependent fractional scatter anchored at $(z=1,f_1)$ and $(z=3,f_3)$:
\begin{equation}
f_{\IGM}(z)=f_1\,z^{-\alpha},\qquad
\alpha=\log_{3}\!\left(\frac{f_1}{f_3}\right),\qquad
\sigma_{\IGM}(z)=f_{\IGM}(z)\,\langle\DM_{\IGM}(z)\rangle .
\end{equation}

\paragraph{Propagation to redshift and comoving distance.}
Using the local slope $\dd\langle\DM\rangle/\dd z\simeq 1000$,
\begin{equation}
\sigma_z(z)\simeq \frac{\sigma_{\DM}(z)}{1000},
\qquad
\sigma_r(z)\simeq \frac{\dd r}{\dd z}\sigma_z=\frac{c}{H(z)}\,\sigma_z.
\label{eq:sigmar}
\end{equation}
We report $\sigma_r$ in $h^{-1}\,\mathrm{Mpc}$ via $\sigma_r[h^{-1}\mathrm{Mpc}]=h\,\sigma_r[\mathrm{Mpc}]$.

\paragraph{Numerical values.}
For the fiducial parameters
($H_0=67.74\,{\rm km\,s^{-1}\,Mpc^{-1}}$, $\Omega_m=0.315$, $h=0.7$,
$f_1=0.13$, $f_3=0.07$,
$\sigma_{\host,0}=50\,{\rm pc\,cm^{-3}}$,
$\sigma_{\rm halo}=30\,{\rm pc\,cm^{-3}}$,
$\sigma_{\ISM}=10\,{\rm pc\,cm^{-3}}$),
we obtain:
\begin{itemize}
\item $z=0.3$: $\sigma_{\rm DM}\simeq 91.6\ {\rm pc\,cm^{-3}}$, \quad $\sigma_r\simeq 241.8\ h^{-1}{\rm Mpc}$.
\item $z=0.5$: $\sigma_{\rm DM}\simeq 106.5\ {\rm pc\,cm^{-3}}$, \quad $\sigma_r\simeq 249.5\ h^{-1}{\rm Mpc}$.
\item $z=0.7$: $\sigma_{\rm DM}\simeq 119.3\ {\rm pc\,cm^{-3}}$, \quad $\sigma_r\simeq 247.4\ h^{-1}{\rm Mpc}$.
\end{itemize}

\section{Simulation and Inference}
\label{sec:pipeline}

To validate our method, we use the public code \texttt{lognormal\_galaxies} to simulate galaxy catalogs at different redshifts.
The input power spectrum is taken to be the matter power spectrum, approximated using the fitting formula of Eisenstein \& Hu,
and is consistent with the Planck 2018 cosmological parameters.
This code enables the generation of mock galaxy catalogs by assuming a lognormal probability density function for both the underlying matter field and the galaxy distribution.

We assume that FRB events occur in a random subsample of galaxies drawn from the simulated galaxy catalog,
which effectively implies $b_{\mathrm{FRB}} = b_{\mathrm{gal}}$.
We generate three sets of galaxy catalogs with linear galaxy bias values $b_{\mathrm{gal}} = [1.2,\ 1.5,\ 2.0]$.
Our estimates of the uncertainties in the comoving distance are based on the calculations presented in Section~\ref{sec:dmz}.
Using the procedure outlined below, we then simulate mock FRB catalogs and test whether the input bias values can be successfully recovered.
For simplicity, we assume that the input galaxy bias is redshift independent; however, our conclusions regarding bias recovery remain unchanged even if an evolving bias model is adopted.

The simulation procedure consists of the following steps:
\begin{enumerate}
    \item At a given redshift, we select a spherical shell with a comoving thickness of $350\,h^{-1}\,\mathrm{Mpc}$.
    This shell thickness is chosen to be sufficiently large such that:
    (i) the estimation of the two-point correlation function does not suffer from spatial incompleteness
    (noting that the Landy--Szalay estimator is, to some extent, insensitive to the survey geometry);
    (ii) a sufficient number of FRB events are contained within the shell; and
    (iii) the true correlation function does not vary significantly across the redshift interval.
    This shell thickness corresponds to a redshift bin width of $\Delta z \simeq 0.3$ at $z=1.0$ and $\Delta z \simeq 0.13$ at $z=0.2$.

    \item From the selected shell, we randomly choose $N$ galaxies to serve as proxies for FRB events.
    For each event, we assign localization uncertainties by assuming a Gaussian posterior distribution (see Section~\ref{sec:method}).

    \item For each of the $N$ points, we draw one realization of the positional offset from the corresponding Gaussian posterior.
    The displaced positions constitute a single realization of an FRB catalog.
    We then measure the two-point correlation function of this Gaussian-smeared catalog using the Landy--Szalay estimator.
    This procedure is repeated $1000$ times, and the resulting correlation functions are averaged to obtain $\xi(r)$ for this set of $N$ events.

    \item To estimate the covariance, we generate $50$ independent galaxy catalogs corresponding to different realizations of the underlying cosmic matter field,
    thereby accounting for cosmic variance.
    For each galaxy catalog, we further construct $20$ random subsamples, each containing $N$ galaxies, to account for sampling variance.
    This yields a total of $1000$ subsamples.
    Each subsample is treated as one realization of the Universe, with its correlation function $\xi(r)$ obtained using the procedure described above.
    The uncertainties on $\xi(r)$ are estimated from the scatter among the remaining subsamples.

    \item The FRB bias factor $b_{\mathrm{FRB}}$ is estimated by comparing the recovered FRB correlation function $\xi_{\mathrm{FRB}}$ with the smeared model correlation function $\xi_{\mathrm{sm}}$.
    The fitting is performed over comoving separations $r \in [10,\,50]\,h^{-1}\,\mathrm{Mpc}$,
    a range that is well matched to the adopted shell thickness and within which the linear bias approximation remains valid.
\end{enumerate}

To quantify the bias recovery, we define a $\chi^2$ statistic,
\begin{equation}
\chi^2 = \sum_{i,j} \Delta X_i \, \Sigma^{-1}_{ij} \, \Delta X_j ,
\end{equation}
where
\begin{equation}
\Delta X_i = \xi_{\mathrm{est}}(r_i) - b^2 \, \xi_{\mathrm{sm}}(r_i) .
\end{equation}
Here, $\xi_{\mathrm{sm}}$ is the (localization-smeared) model correlation function estimated from the simulated catalogs in each radial bin,
and $\Sigma_{ij}$ denotes the covariance matrix between the $i$-th and $j$-th bins.
The covariance matrix is estimated using the $1000$ subsamples described above.

We further define a likelihood function and perform a Bayesian analysis to estimate the posterior distribution of the bias parameter $b$.
A uniform prior is assumed over the range $b \in [0,\,5]$.
The posterior sampling is carried out using the nested sampling algorithm implemented in the \texttt{DYNESTY} package.

Since galaxies are used as proxies for FRB events, we expect the recovered $b_{\mathrm{FRB}}$ to be consistent with the input bias parameter used in the simulated galaxy distribution.
Owing to the relative insensitivity of the Landy--Szalay estimator to survey geometry.

The marginalized $68\%$ credible-region widths of the posterior distributions in right ascension, declination, and comoving distance are modeled using a multivariate Gaussian distribution centered at zero, with characteristic standard deviations of $(\sigma_{\mathrm{RA}},\, \sigma_{\mathrm{Dec}},\, \sigma_{r}) = (0.5,\,0.5,\,240)$.

For the angular components, we adopt truncated Gaussian uncertainties. Based on our mock localization setup, angular uncertainties under the degree level do not introduce a significant impact on the two-point correlation function in comoving distance.

The corresponding comoving distance uncertainties are computed following the numerical values described in the previous section.
Correlations among right ascension, declination, and comoving distance uncertainties are neglected.

\section{Results}
\label{sec:results}

In this section we present the recovery of the FRB clustering signal and the corresponding constraints on the linear bias parameter $b$ obtained from the two-point correlation function (2PCF) measured from mock FRB catalogs with localization uncertainties.

\begin{figure}[t]
    \centering
    \includegraphics[width=0.92\linewidth]{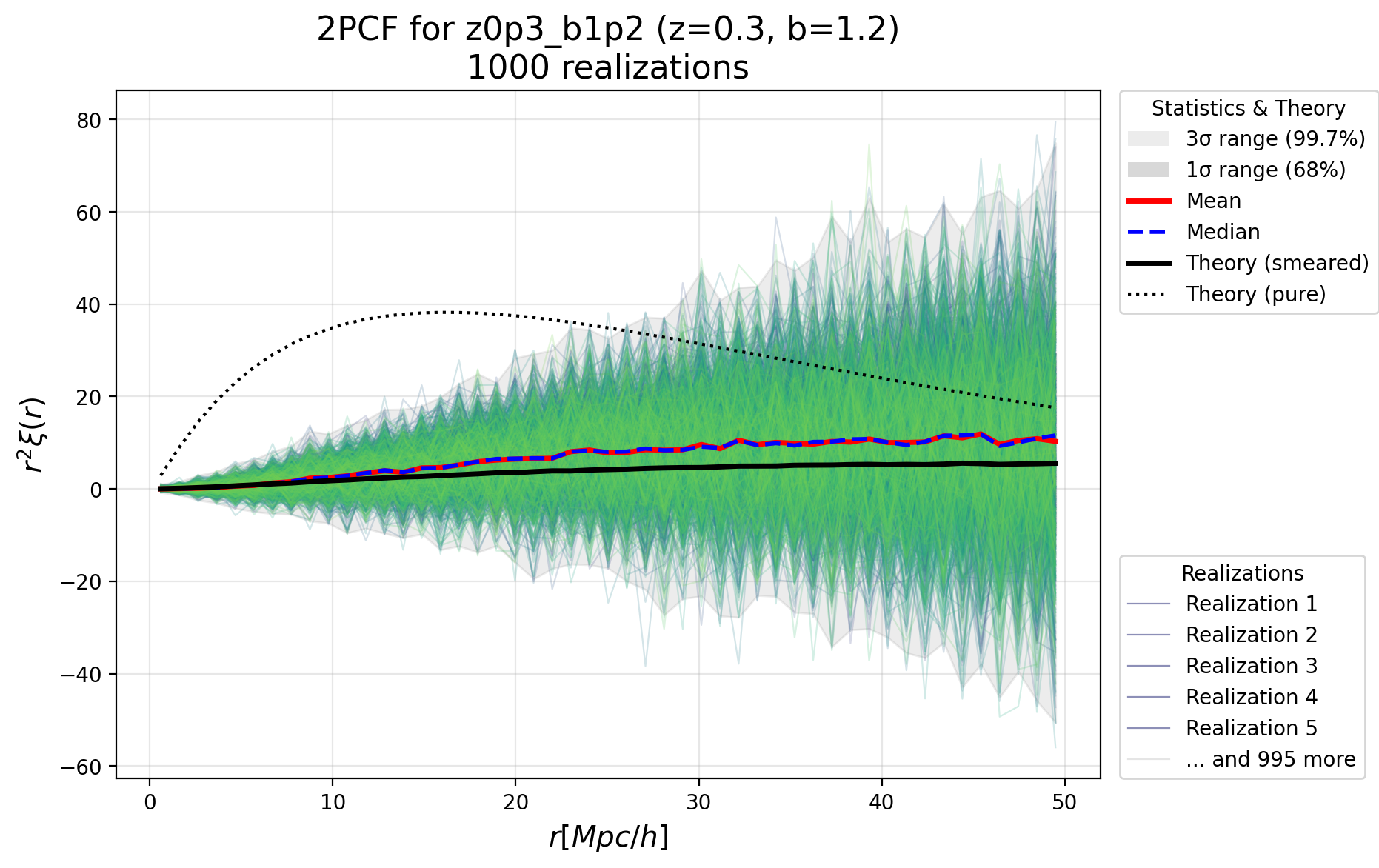}
    \caption{Representative recovery of the localization-smeared two-point correlation function for the mock
    sample with $z=0.3$ and $b_{\rm inj}=1.2$. The dotted black curve shows the unsmeared theoretical
    prediction, and the solid black curve shows the smeared prediction after convolving with the adopted
    localization kernel. Thin colored curves denote the individual measurements from 1000 realizations; the red
    solid and blue dashed curves mark the ensemble mean and median, respectively. The dark and light gray
    shaded regions indicate the $1\sigma$ and $3\sigma$ ranges.}
    \label{fig:xi_example}
\end{figure}

\paragraph{Recovery of the localization-smeared 2PCF.}
Figure~\ref{fig:xi_example} shows a representative example of the recovered statistic $r^2\xi(r)$ for the $z=0.3$ and $b_{\rm inj}=1.2$ mock sample. In this figure, the dotted black curve denotes the unsmeared theoretical 2PCF, while the solid black curve shows the corresponding localization-smeared prediction after propagating the FRB positional uncertainties through the forward model described in Sections~\ref{sec:method} and \ref{sec:pipeline}. The thin colored curves represent the individual measurements from 1000 realizations, the red solid curve marks the mean, the blue dashed curve marks the median, and the gray bands indicate the $1\sigma$ and $3\sigma$ ranges.

Two features are immediately apparent. First, the measured mean and median track the smeared theoretical prediction rather than the unsmeared one, showing that the dominant effect of localization and DM-based radial uncertainties is to suppress the observable clustering amplitude in comoving space. Second, the realization-to-realization scatter is substantial, especially toward larger separations, indicating that the measurement remains noise dominated for the current mock configuration. Nevertheless, the injected clustering level still falls within the statistical range recovered from the ensemble, which shows that the pipeline is able to recover the FRB bias at the correct order of magnitude once the smearing is modeled consistently.

\begin{figure*}[t]
\centering
\begin{minipage}{0.32\textwidth}
    \centering
    \includegraphics[width=\linewidth]{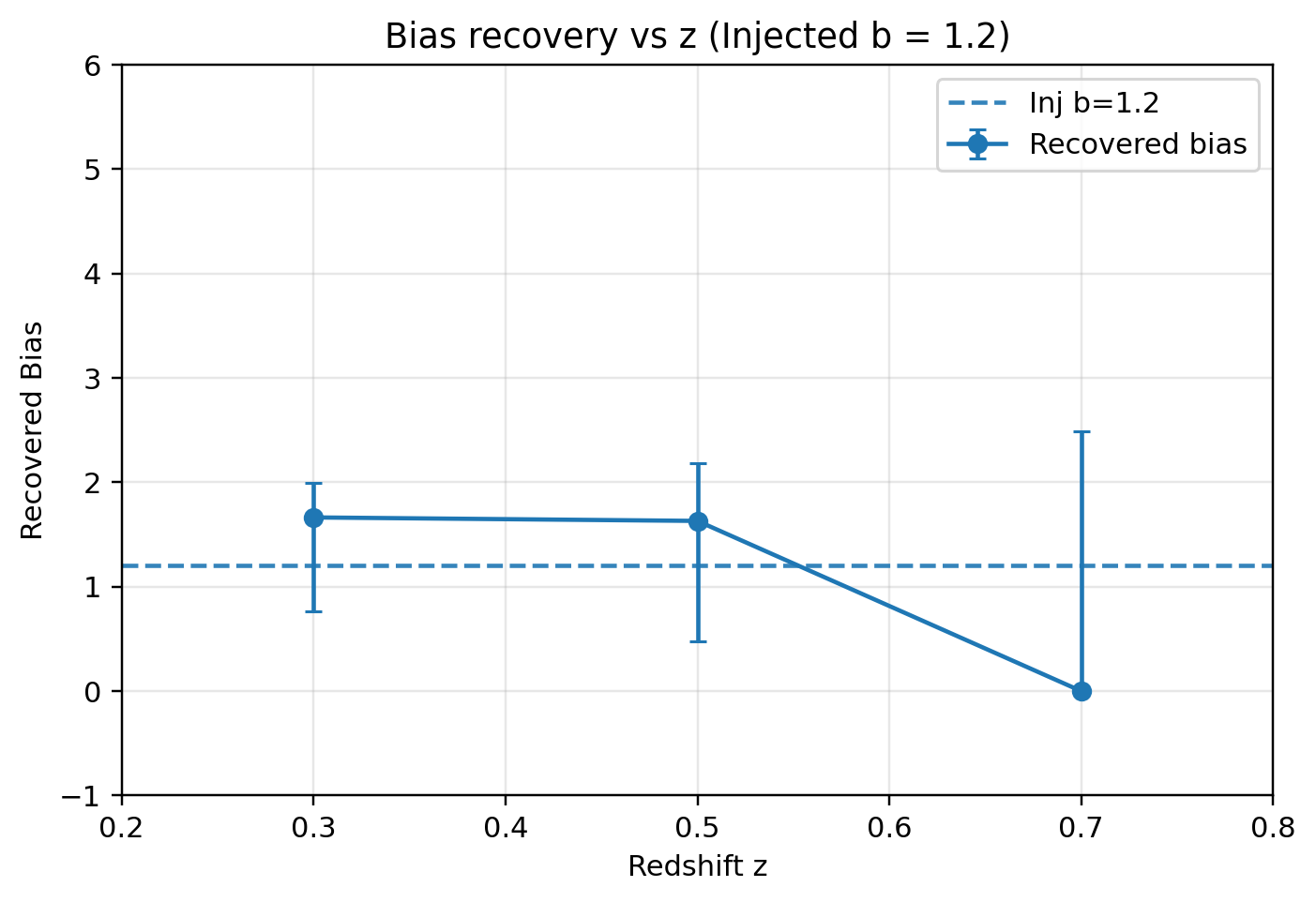}
\end{minipage}\hfill
\begin{minipage}{0.32\textwidth}
    \centering
    \includegraphics[width=\linewidth]{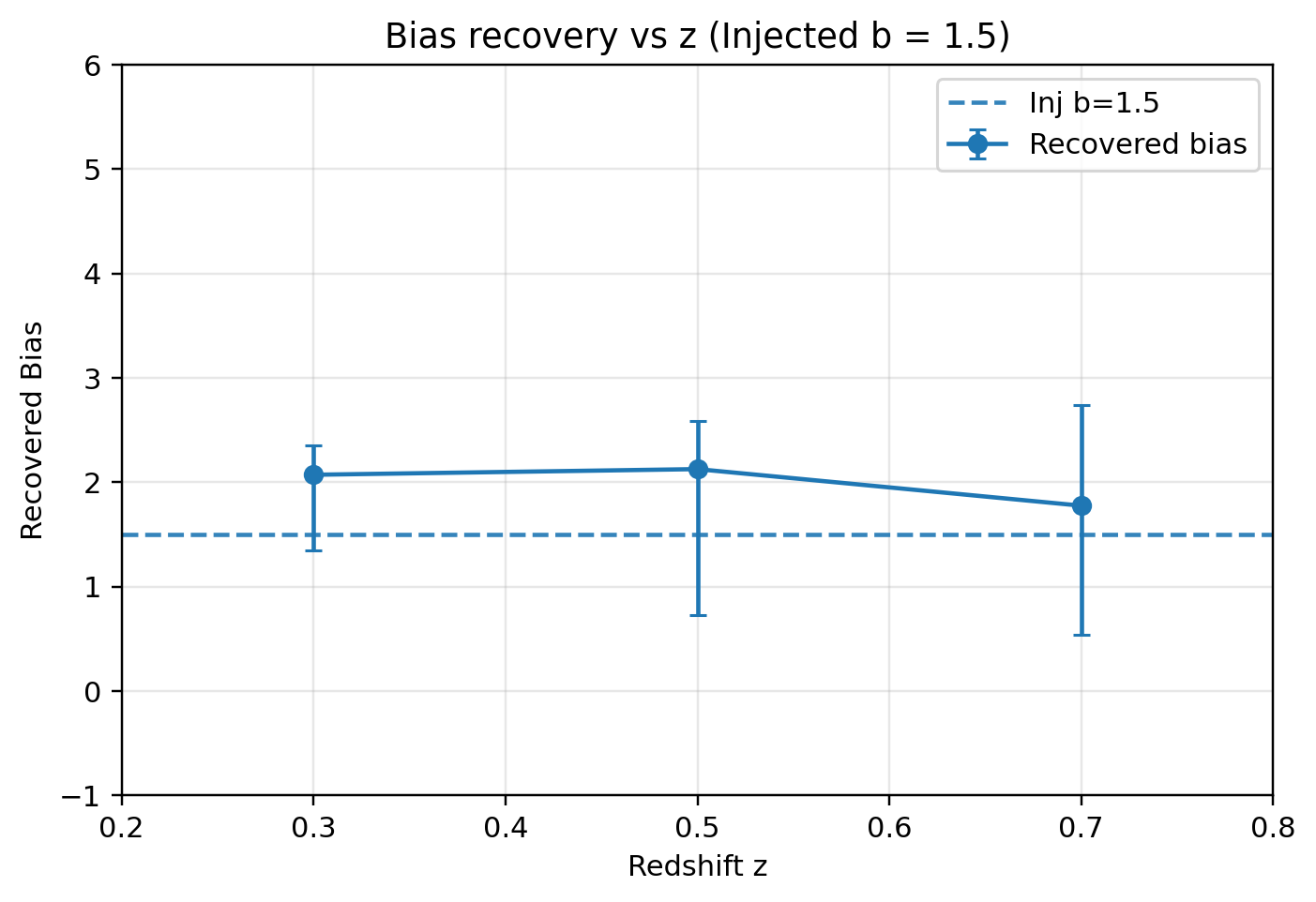}
\end{minipage}\hfill
\begin{minipage}{0.32\textwidth}
    \centering
    \includegraphics[width=\linewidth]{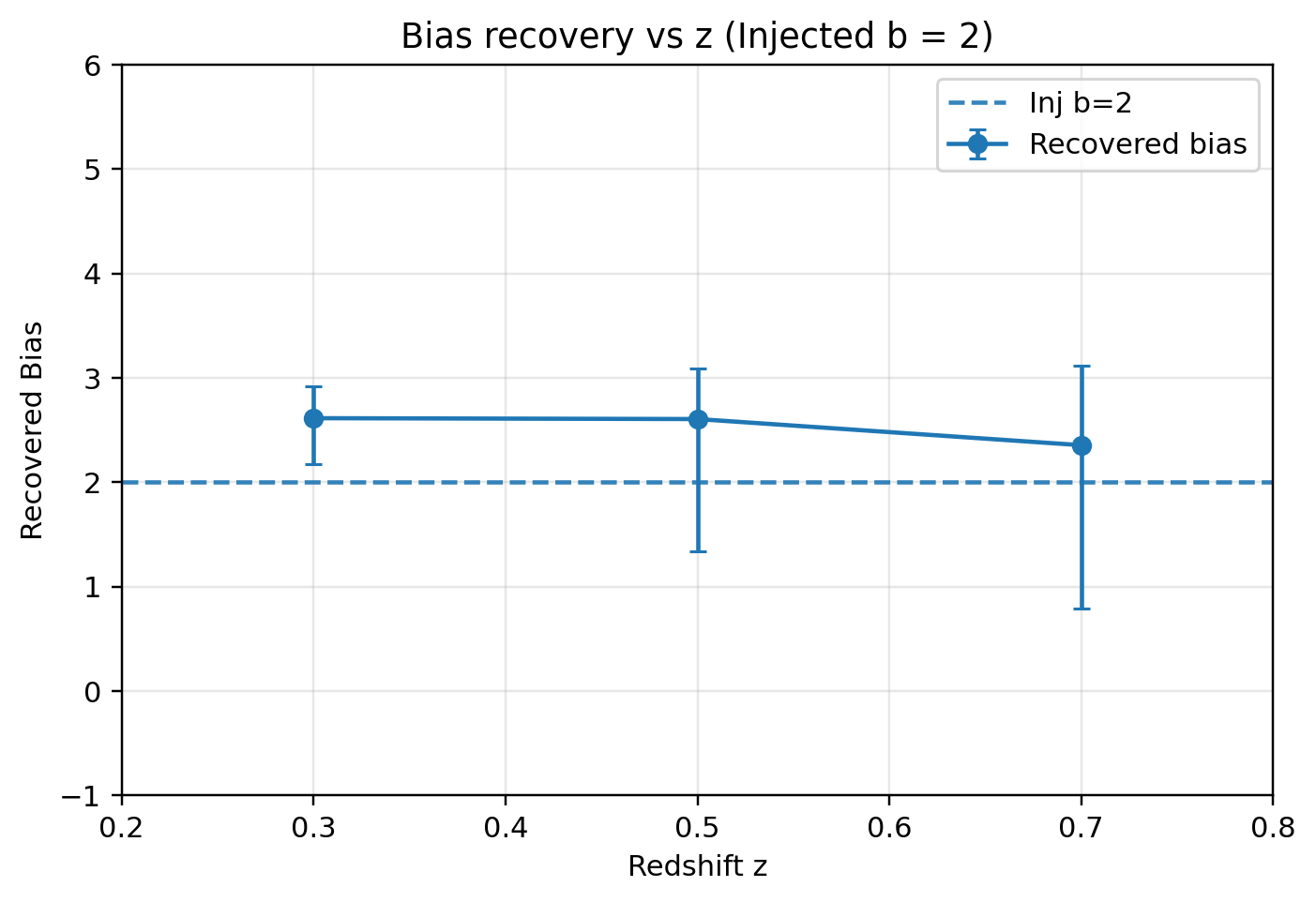}
\end{minipage}
\caption{Recovered FRB bias as a function of redshift for three injected values,
$b_{\rm inj}=1.2$, $1.5$, and $2.0$ (from left to right). The dashed horizontal line in each panel marks the
injected bias. Points with vertical error bars show the recovered posterior central value and the corresponding
$68\%$ credible interval in each redshift bin. The overall trend indicates that the constraining power weakens
toward higher redshift, especially for the lowest-bias case.}
\label{fig:bias_vs_z}
\end{figure*}

\paragraph{Bias recovery as a function of redshift.}
Figure~\ref{fig:bias_vs_z} summarizes the recovered FRB bias as a function of redshift for three injected values, $b_{\rm inj}=1.2$, $1.5$, and $2.0$. Each point gives the recovered central value from the posterior, and the vertical bars indicate the corresponding $68\%$ credible interval. The dashed horizontal line in each panel marks the injected bias.

Overall, the method recovers the input bias reasonably well across most redshift bins, with two notable exceptions. Owing to the intrinsically weak clustering at high redshift and the further suppression from spatial smearing, the low-bias case $b_{\rm inj}=1.2$ loses effective constraining power at $z=0.7$. Meanwhile, the recovered bias is slightly overestimated at $z=0.3$ for $b_{\rm inj}=2.0$, though the deviation remains moderate. For $b_{\rm inj}=1.5$ and $2.0$, central values stay close to injected values with strong constraints from $z=0.3$ to $z=0.7$. For $b_{\rm inj}=1.2$, central values at $z=0.3$ and $0.5$ are slightly elevated (but consistent with the posterior), while constraints weaken drastically at $z=0.7$, approaching the limit for distinguishing low-bias FRB populations from the prior.

\begin{figure*}[t]
\centering
\begin{minipage}{0.32\textwidth}
    \centering
    \includegraphics[width=\linewidth]{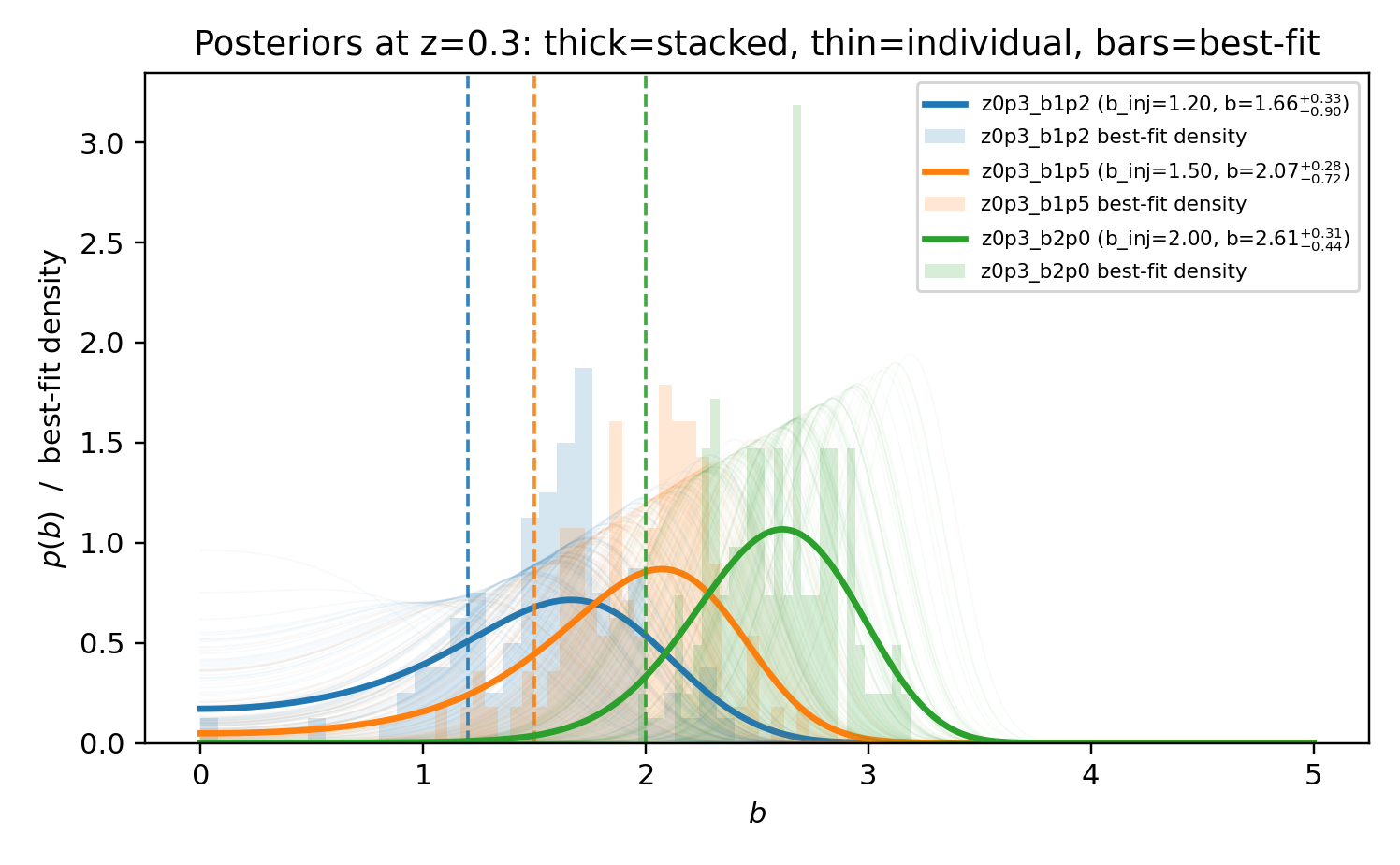}
\end{minipage}\hfill
\begin{minipage}{0.32\textwidth}
    \centering
    \includegraphics[width=\linewidth]{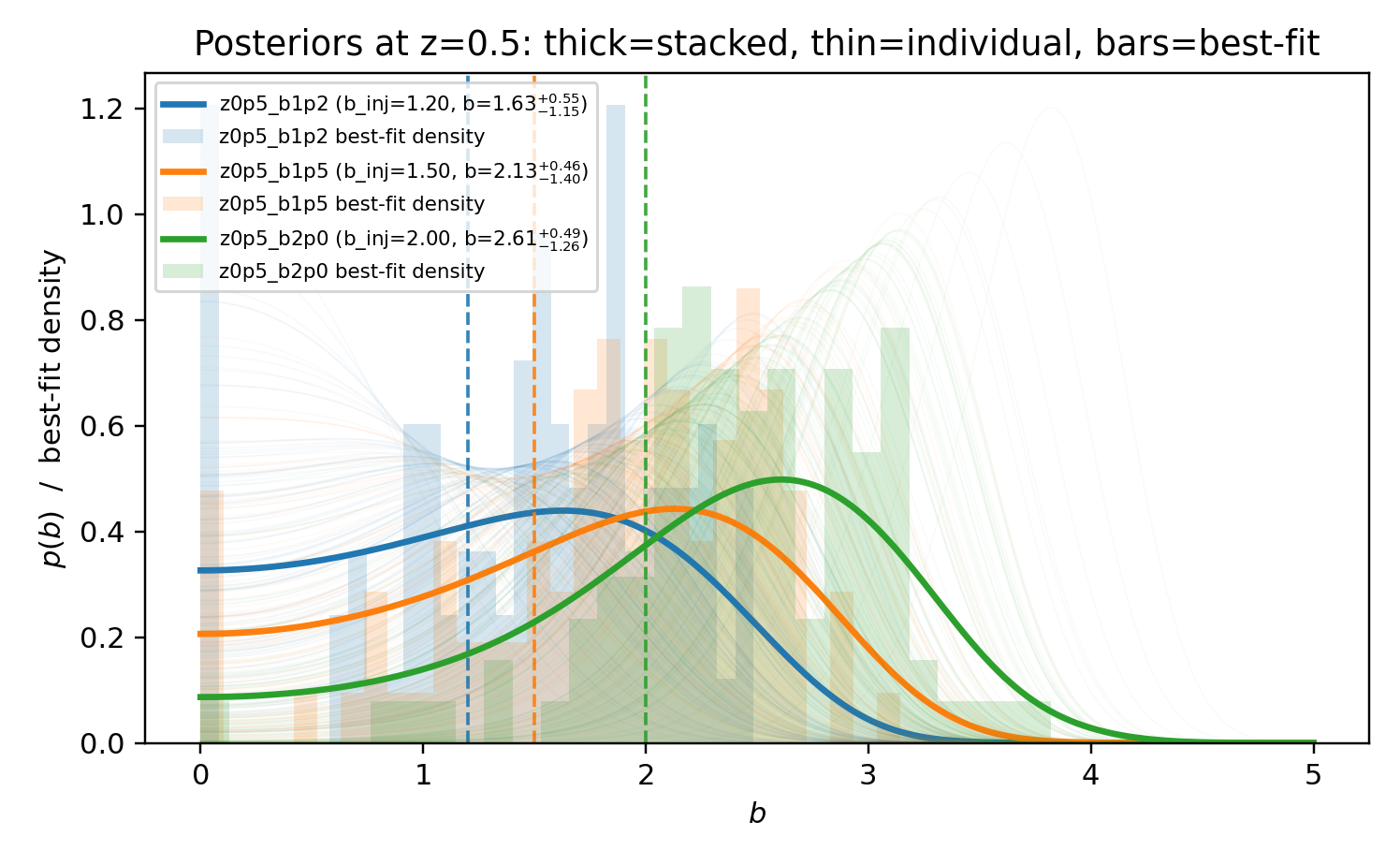}
\end{minipage}\hfill
\begin{minipage}{0.32\textwidth}
    \centering
    \includegraphics[width=\linewidth]{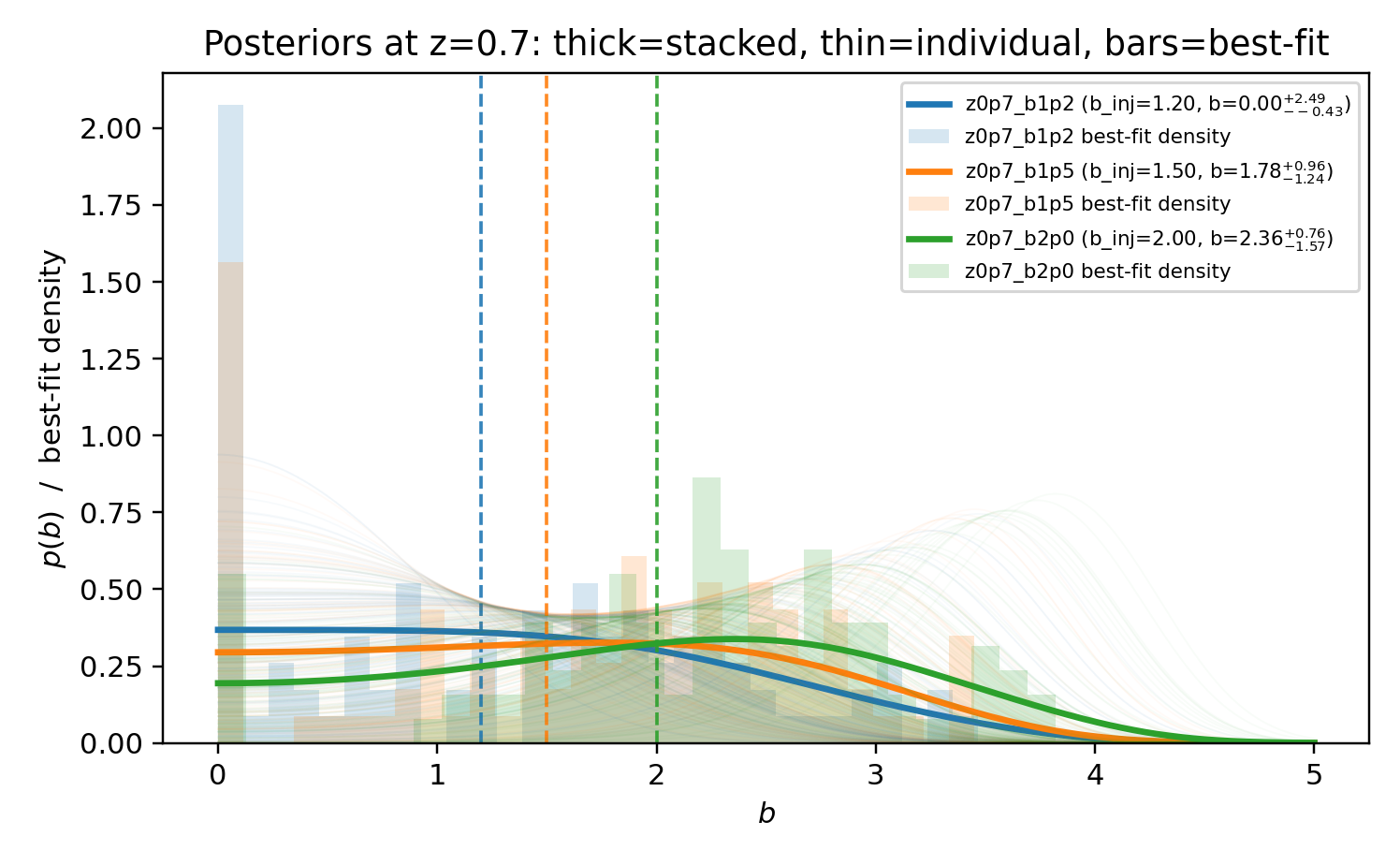}
\end{minipage}
\caption{Stacked posterior densities of the FRB bias parameter $b$ in the three redshift bins
$z=0.3$, $0.5$, and $0.7$. Blue, orange, and green correspond to injected biases
$b_{\rm inj}=1.2$, $1.5$, and $2.0$, respectively. Thick curves denote the stacked posterior densities, thin curves
show the posteriors from individual realizations, semi-transparent histograms indicate the best-fit-density
distribution, and vertical dashed lines mark the injected values. The separation between the three bias cases is
clearest at low redshift and weakens toward $z=0.7$, especially for the lowest-bias sample.}
\label{fig:stacked_posteriors}
\end{figure*}

\paragraph{Stacked posteriors and global separation of different bias hypotheses.}
Figure~\ref{fig:stacked_posteriors} provides a complementary summary of the bias inference by stacking the posterior information from the 1000 realizations in each redshift bin. In each panel, the thick curves show the stacked posterior density for the three injected bias cases, the thin curves show the individual posteriors from single realizations, the semi-transparent histograms represent the distribution of best-fit values, and the vertical dashed lines mark the injected biases.

This figure is useful because it visualizes not only the typical posterior width, but also the degree of overlap between different injected-bias populations. At $z=0.3$, the three stacked posteriors are clearly ordered and remain reasonably well separated, indicating that the method can statistically distinguish FRB samples with different large-scale clustering amplitudes in the low-redshift regime. At $z=0.5$, the posteriors become broader and the overlap increases, but the ranking with injected bias is still preserved and the three cases remain identifiable in a statistical sense. By $z=0.7$, the degradation becomes more obvious: the posterior for $b_{\rm inj}=1.2$ broadens strongly and develops substantial support near the lower prior boundary, whereas the $b_{\rm inj}=1.5$ and $2.0$ cases still show identifiable but significantly weaker separation.

\section{Discussion}
\label{sec:discussion}

The strong suppression of the 2PCF amplitude relative to the pure theory curve is physically expected. In the present analysis, the dominant radial uncertainty is inherited from the full $\DM \rightarrow z \rightarrow r$ propagation chain. Because the total DM variance receives non-negligible contributions from several components---including the host term, Galactic foregrounds, and the IGM scatter---the effective comoving-distance uncertainty remains large, reaching the level of a few hundred $h^{-1}\,{\rm Mpc}$. Such a broad radial kernel smooths the underlying three-dimensional point distribution, thereby lowering the measured 2PCF amplitude and weakening the visible peak structure of the correlation function. At the same time, when only $\sim 10^4$ FRBs are available, this level of positional uncertainty also significantly enlarges the measurement variance, which is directly reflected in the broad error bands in Figure~\ref{fig:xi_example}.

The growth of the error bars with redshift in Figure~\ref{fig:bias_vs_z} has two main causes. First, the underlying matter 2PCF becomes intrinsically less prominent at higher redshift in the radial range used for the fit, so the same level of observational smoothing produces a larger fractional loss of information. Second, in the present mocks the total number of FRBs is fixed in each redshift bin, while the comoving volume represented by a higher-redshift shell is larger. As a result, the effective FRB number density decreases toward high redshift, which increases the shot noise and directly broadens the posterior on $b$. Importantly, this deterioration is not driven primarily by a dramatic increase in the FRB radial smearing itself. In the $\DM$--$z$--distance chain adopted here, the effective comoving-distance uncertainty does not differ strongly between low and high redshift. At low redshift, although the total extragalactic DM is smaller, the fractional importance of the non-IGM terms---such as the host and Galactic contributions---is larger, so the final radial uncertainty is not especially small. At higher redshift, the IGM contribution becomes dominant and dilutes the relative impact of those other components. Consequently, the net comoving-distance uncertainty remains of comparable order across the three redshift bins.

Taken together, Figures~\ref{fig:bias_vs_z} and \ref{fig:stacked_posteriors} suggest that the present pipeline is already able to recover the approximate FRB bias scale and to preserve the ordering of different bias models, but that the inference is not yet in a precision regime. In particular, the stacked posteriors show a mild tendency for the recovered central values to lie somewhat above the injected ones at $z=0.3$ and $z=0.5$, which likely indicates that subdominant calibration effects in the current simplified forecasting setup are still non-negligible. For the present stage of the work, this does not invalidate the main conclusion---namely, that a forward-modeled localization-smearing treatment allows useful bias inference from unlocalized FRB samples---but it does show that a refined treatment of the DM uncertainty budget and survey configuration will be important in future applications.

More generally, the present analysis should be viewed as one specific realization of an FRB clustering forecast in reconstructed comoving space. Future FRB surveys will provide larger samples, including both localized and unlocalized events, which should improve the statistical control of the covariance and allow more informative bias constraints. At the same time, improved localization capabilities and a higher fraction of host-identified FRBs should tighten the constraints on the separate DM components and reduce the uncertainty of the $\DM$--$z$--distance mapping itself. FRB large-scale structure studies also need not be restricted to the comoving-space approach adopted here; complementary analyses in DM space or cross-correlations with galaxy samples may provide additional constraining power \citep{RafieiRavandi2021LSS,SagaAlonso2024DMspace,Hussaini2025DMLSS}.

\begin{acknowledgments}
This work was supported by the National Key Research and Development Program of China Grant Nos. 2023YFC2206702, and 2021YFC2203001; National Natural Science Foundation of China under Grants Nos.11920101003, 12021003, 11633001, 12322301, and 12275021; the Strategic Priority Research Program of the Chinese Academy of Sciences, Grant Nos. XDB2300000 and the Interdiscipline Research Funds of Beijing Normal University.
\end{acknowledgments}

\bibliography{ref}{}
\bibliographystyle{aasjournalv7}

\end{document}